\documentclass[twocolumn]{aa}

\usepackage[T1]{fontenc}
\usepackage{ae,aecompl}
\usepackage{lipsum}
\usepackage{lscape}
\usepackage{natbib}
\usepackage{listings}

\usepackage{realboxes}
\bibpunct{(}{)}{;}{a}{}{,} 

\usepackage{graphicx}	
\usepackage{amsmath}	
\usepackage{amssymb}	
\usepackage[dvipsnames]{xcolor}
\usepackage{txfonts}
\usepackage[colorlinks=true,allcolors=cyan]{hyperref}
\usepackage{ulem}
\usepackage{multirow}

\defcitealias{zhang2024hot}{Z24}
\defcitealias{aird2013primus}{A13}
\defcitealias{bekhti2016hi4pi}{HI4PI collaboration et al. 2016}
\defcitealias{ade2016planck}{Planck Collaboration XIII 2016}
\defcitealias{anderson2015unifying}{A15}
\newcommand{\MSUN}{{M}_\odot}
\newcommand{\Mtwo}{{M}_{\rm 200m}}
\newcommand{\Rtwo}{{ R}_{\rm 200m}}

\begin{document} 

   \title{Impact of the large-scale cosmic web on the X-ray emitting circumgalactic medium}

    \author{Soumya Shreeram\inst{1}\thanks{\href{shreeram@mpe.mpg.de}{shreeram@mpe.mpg.de}}, Daniela Gal\'arraga-Espinosa\inst{2}, Johan Comparat\inst{1}, Andrea Merloni\inst{1},  Daisuke Nagai\inst{3, 4}, C\'eline Peroux\inst{5}, Ilaria Marini\inst{5}, C\'eline Gouin\inst{6}, Kirpal Nandra\inst{1}, Yi Zhang\inst{1}, Gabriele Ponti\inst{1, 7}, and Anna Olechowska\inst{1} }

    \institute{Max Planck Institute for Extraterrestrial Physics (MPE), Gie\ss enbachstraße 1, 85748 Garching, Munich, Germany
    \and
        Kavli IPMU (WPI), UTIAS, The University of Tokyo, Kashiwa, Chiba 277-8583, Japan\label{IPMU}
    \and
        Department of Physics, Yale University, New Haven, CT 06520, USA
    \and    
        {Department of Astronomy, Yale University, New Haven, CT 06520, USA}
    \and 
        European Southern Observatory, Karl-Schwarzschild-Stra\ss e 2, 85748 Garching, Munich, Germany
    \and
        Institut d'Astrophysique de Paris, 98bis Bd Arago, 75014 Paris, France
    \and
        INAF-Osservatorio Astronomico di Brera, Via E. Bianchi 46, I-23807 Merate (LC), Italy
        }
    \date{Received ZZZZ, ZZZZ; accepted XXXX, XXXX; published YYYY, YYYY}

\abstract
  {}
  {The hot circumgalactic medium (CGM), probed by  X-ray observations, plays a central role in understanding gas flows that drive a galaxy's evolution. While CGM properties have been widely studied, the influence of a galaxy's large-scale cosmic environment on the hot gas content remains less explored. We investigate how the large-scale cosmic web affects the X-ray surface brightness (XSB) profiles of galaxies in the context of cosmological simulations.}
  {We use our novel IllustrisTNG-based lightcone, spanning $0.03\leq z\leq0.3$, first developed in our previous work, and generate self-consistent mock X-ray observations, using intrinsic gas cell information. We apply the filament finder DisPerSE on the galaxy distributions to identify the cosmic filaments within the lightcone. We classify central galaxies into five distinct large-scale environment (LSE) categories: clusters and massive groups, cluster outskirts, filaments, filament-void transition regions, and voids/walls.
  }
  {We find that the X-ray surface brightness profiles (XSB) of central galaxies of dark matter haloes in filaments with $\Mtwo >10^{12}\ \MSUN$ are X-ray brighter than those in voids and walls, with $20-45\%$ deviations in the radial range of $(0.3-0.5)\times \Rtwo$. We investigate the source of this enhancement and find that the filament galaxies show higher average gas densities, temperatures, and metallicities compared to voids/walls galaxies. }
  {Our results demonstrate that the impact of the large-scale cosmic environment is imprinted on the hot CGM's X-ray emission. Future theoretical work on studying the effect of assembly history, connectivity, and gas accretion on galaxies in filaments and voids would help to further our understanding of the impact of the environment on X-ray observations.}
    \titlerunning{Effect of the Large Scale Structure on the Hot CGM in X-rays}
    \authorrunning{S. Shreeram}
   \maketitle


\section{Introduction}

Observational and theoretical studies show that galaxy properties, such as stellar mass, star formation rate, and gas content, vary across different large-scale environments (LSE). The galaxies close to groups and clusters, which form the nodes of the cosmic web,  are more likely to be elliptical, red, and have suppressed star formation, compared to their less crowded “field” counterparts that tend to be spiral, blue, and actively forming stars \citep{dressler1980galaxy, butcher1984evolution, dressler1997evolution, lewis20022df, blanton2005relationship, alpaslan2015galaxy, pasquali2015environment, shimakawa2021subaru}. 
Additionally, galaxies infalling into clusters via cosmic filaments are systematically more quenched than their counterparts from other isotropic directions (see e.g. \citealt{martinez2016galaxies, einasto2018supercluster, salerno2019filaments, gouin2020probing}). Simulations show that the gas content of galaxies located as far as within $3\times$ virial radius~\citep{cen2014gas, arthur2019thethreehundred, mostoghiu2021three} to $5\times$ the virial radius~\citep{bahe2013does} of the groups and clusters centre is gas-depleted compared to their counterparts in the field, as also reaffirmed by observations  (e.g., \citealt{tanaka2004environmental, catinella2013galex, cortese2011effect}). Similar trends of higher gas depletion, higher quiescent fraction, and stellar mass also hold for galaxies closer to the cosmic filament spines~\citep{Malavasi2017vimos, laigle2018cosmos2015, sarron2019pre, bonjean2020filament, winkel2021imprint, hoosain2024effect}.

The hot CGM, which is the diffuse gas that surrounds galaxies, plays a crucial role in regulating the growth and evolution of the galaxy~(see \citealt{tumlinson2017circumgalactic} and \citealt{faucher2023key} for a review). The hot gas ($T\gtrsim 10^6$ K) reservoir hosted by the CGM is crucial for replenishing the cold gas consumed for star formation~\citep{fox2017gasacc, wang2022large}. The ability of the halos to retain or deplete their cold gas in an intrafilamentary environment has been shown to be correlated with the stellar mass in HI studies, see e.g.,~\citealt{kleiner2017evidence, odekon2018effect, hoosain2024effect}. From these HI  studies, the emerging picture is that mass plays a crucial role in determining whether a galaxy can retain/further accrete gas from the cosmic web, or is more vulnerable to stripping and supply truncation. More massive galaxies ($M_\star>10^{11}\ \MSUN$) retain and accrete the gas from the surrounding filament due to their deeper gravitational potentials; however, the lower-mass ($M_\star<10^{10.5}\ \MSUN$) galaxies are more subject to gas-depleting processes like stripping and detachment. However, there is a lack of understanding of how the hot gas content around galaxies, as probed by the CGM, is impacted by filaments, voids, and nodes. This motivates the need to explore whether the hot CGM, as traced through X-ray emission, encodes information about a galaxy’s cosmic web environment.

In the $\Lambda$CDM Universe, the existence of the cosmic web follows from the initial fluctuations in the primordial density field, whose evolution is dictated by gravity in an expanding Universe. The anisotropic nature of gravitational collapse leads to the formation of high-density peaks that are the nodes that host today's galaxy clusters, and the expansive network of bridges between these nodes forms a large-scale web dominated by filaments, which demarcate the underdense voids~\citep{peebles2020large}. The theoretical formulation of the existence and evolution of the cosmic web~\citep{bond1996filaments} has been confirmed by all large N-body simulations of structure formation in a $\Lambda$CDM Universe (e.g. \citealt{klypin1983three, springel2006large, Popping2009h1, Angulo2012scaling, Habib2012largesim, Poole2015wiggleZ}). The existence of filaments, clusters and voids is also reaffirmed with advances in spectroscopic surveys, with increasing resolution and depth, which have allowed us to observationally map the cosmic web. A unified approach to jointly study cosmic web, as traced by galaxies, is possible with exquisite detail up to redshift z $\approx$ 0.9 with surveys such as the CfA Redshift Survey~\citep{de1986slice}, SDSS~\citep{york2000sloan}, 2dFGRS~\citep{colless20012df}, 6dFGS~\citep{jones20096df}, GAMA~\cite{driver2011galaxy}, Vipers~\citep{guzzo2014vimos}, 2MASS~\citep{huchra20122mass} and COSMOS~\citep{scoville2007large}. This is being further pushed to higher redshifts of about z $\approx$ 2, close to the peak epoch of star formation with ongoing and upcoming stage-4 surveys such as Euclid~\citep{laureijs2011euclid}, PFS~\citep{takada2014extragalactic}, 4MOST~\citep{de20124most}.

Studies using simulations find that cosmic filaments dominate the mass budget, occupying $50\%$ of the total mass of the cosmic web, with mean overdensities $\delta\sim 10$~\citep{cautun2014evolution}, while the underdense voids, $\delta\sim -0.8$, are the most voluminous component of the cosmic web~\citep{sheth2004hierarchy}. \cite{cui2018large} show that the gas component is the dominant baryonic tracer of cosmic filaments, hosting the warm-hot intergalactic medium (WHIM) gas phase~\citep{galarraga2021properties}. The WHIM gas can be accreted onto the halos resulting in the denser circumgalactic medium (CGM) gas phase ($n_{\rm H} \gtrsim 10^{-4}$ cm$^{-3}$; see categorisation in \citealt{martizzi2019baryons}), and inversely, CGM gas around halos might be ejected due to feedback effects or undergo stripping due to ram-pressure inside filaments~\citep{benitez2013dwarf, winkel2021imprint}. \cite{liao2019impact} show that up to $30\%$ of the gas accreting onto galaxies residing in filaments is pre-processed, and they also have higher baryon fractions compared to the field galaxies~(e.g., see also \citealt{singh2020study}). The CGM, as probed in X-rays, is thus an interesting avenue to test for these environment-driven gas processes. 

In this work, we use an IllustrisTNG-based lightcone from \cite{shreeram2025quantifying, shreeram2025retrieving}, LC-TNGX, to study the impact of the LSE on the hot gas properties of galaxies. Isolating the impact of the large-scale environment (LSE) is complicated by the fact that various mechanisms, both gravitational and hydrodynamic, act simultaneously on galaxies. Particularly, local overdensity and galaxy hierarchy (central vs satellite) also impact galaxy properties~\citep{pasquali2015environment, o2024effect, rodriguez2024evolutionary}. The latter effect can be accounted for by separately studying central/satellite galaxy trends of galaxy properties~\citep{yu2025impact}. The former effect of local overdensity (or crowdedness of the environment) is related to different cosmic web environments having degeneracies between their local and global overdensities~\citep{hahn2007properties, cautun2014evolution, o2024effect}. As the local overdensity correlates with the halo mass function, where massive halos reside in high-density regions \citep{Tinker2011halo_lse, wang2018dearth, Wechsler2018halo}, in this work, we account for the local density effects by studying the impact of the LSE in halo mass bins. In this way, we additionally also distinguish the effect of the stellar-to-halo-mass relation (SHMR) from that of the LSE (e.g, see \citealt{Wechsler2018halo}).

The paper is organised as follows. We describe  the LC-TNGX and how we self-consistently generate mock X-ray observations within the LC-TNGX, using the gas cell information in Sec.~\ref{subsec:lc_tng300}. We apply DisPerSe on the galaxy distribution to identify the cosmic filaments within LC-TNGX, as detailed in Sec.~\ref{subsec:disperse}. We classify the central galaxies in different LSE: clusters and groups, galaxies in clusters and group outskirts, galaxies in filaments, galaxies in the filament-void transition region, and galaxies in voids and walls in Sec.~\ref{subsec:classification}. Sec.~\ref{sec:results} presents the main results of this work on how the XSB profiles are affected by the galaxies in different LSE. We interpret our findings in Sec.~\ref{sec:discussion} and report our conclusions in Sec.~\ref{sec:conclusions}.

 \begin{figure*}
    \centering
    \includegraphics[width=\textwidth]{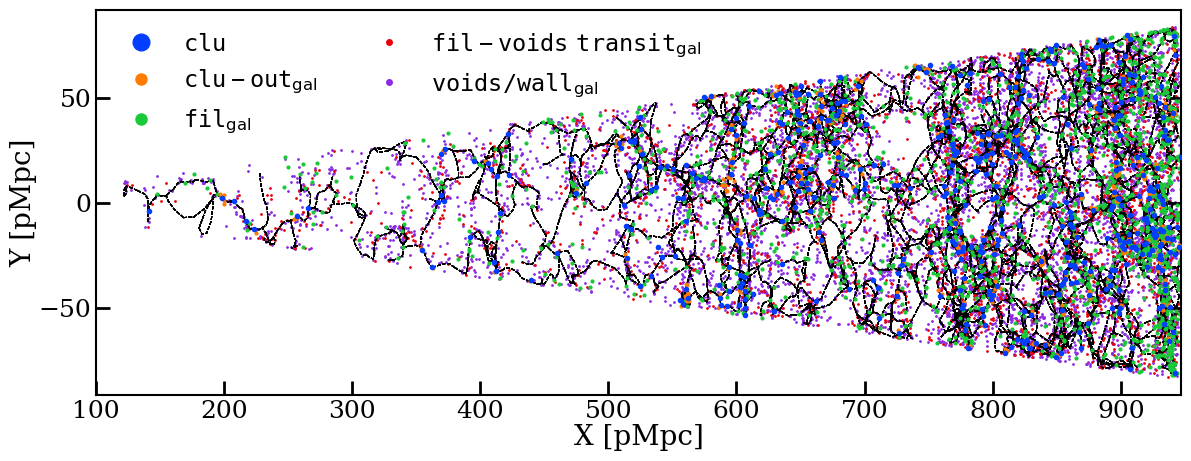}
    \includegraphics[width=\textwidth]{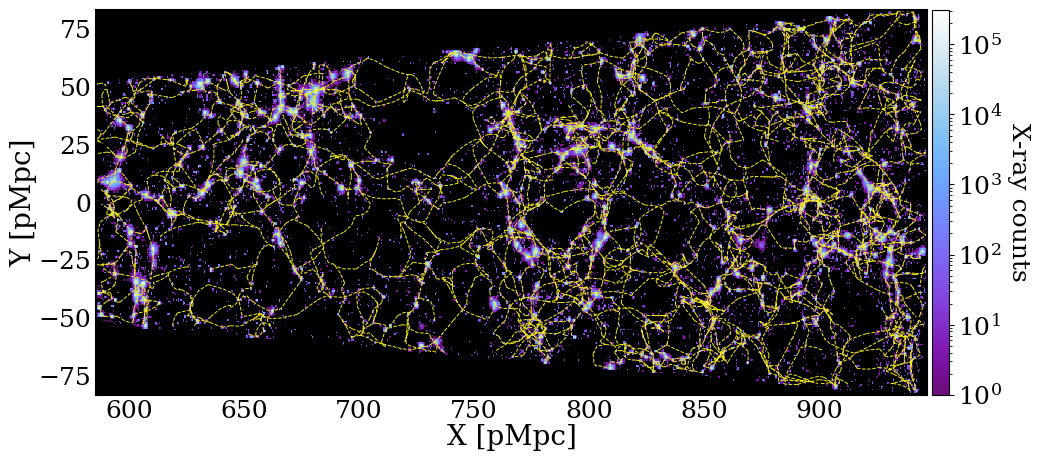}
    \caption{The TNG300 lightcone (LC-TNGX) built in \cite{shreeram2025quantifying} overplotted with the filaments, which are identified using the DisPerSe algorithm (Sec .~\ref {subsec:disperse}). The top panel shows the central galaxies in LC-TNGX that are classified into five distinct large-scale environments categories: (1) clusters in blue, (2) cluster outskirts in orange, (3) filaments in green, (4) filament-void transition region in red, and (5) voids/walls in purple, defined in Sec.~\ref{subsec:classification}. The bottom panel zooms into the lightcone, showing the X-ray events generated with \texttt{pyXsim} in the $0.5-2.0$ keV energy band with the filaments depicted by the yellow-dashed lines. While the filaments are identified using the galaxy distribution, we show that the hot gas emitting X-ray events trace the cosmic web. This alignment validates the full pipeline used in this work, from the lightcone construction and X-ray map generation to filament identification.}
    \label{fig:lss}
\end{figure*}
\section{Methods}
\label{sec:methods}

This section details the data products used for studying the impact of the LSE on the hot CGM. More precisely, Sec.~\ref{subsec:lc_tng300} describes the TNG300 X-ray lightcone (LC-TNGX), Sec.~\ref{subsec:disperse} describes the filament catalogue obtained within the LC-TNGX using DisPerSE, and lastly, Sec.~\ref{subsec:classification} describes the classifications of the LC-TNGX halos into their LSE categories.
\subsection{The TNG300 X-ray lightcone: LC-TNGX}
\label{subsec:lc_tng300}

In this work, we model the hot gas emission using the TNG300 hydrodynamical simulations \citep{pillepich2018first, marinacci2018first, naiman2018first, nelson2015illustris, springel2018first}; we use TNG300\footnote{\url{ http://www.tng-project.org} } to construct a lightcone, using the \texttt{LightGen} code\footnote{The code to generate ligthcones from TNG is publically available at \url{https://github.com/SoumyaShreeram/LightGen/}}, and generate mock X-ray observations (LC-TNGX), which is presented in~\cite{shreeram2025quantifying} and applied to eROSITA data in \cite{shreeram2025retrieving}. Here, we summarise the most important features. Using the IllustrisTNG cosmological hydrodynamical simulation, with the box of side length $302.6$ Mpc~\citep[TNG300]{nelson2019illustristng}, we map the hot CGM around a wide range of halo masses embedded in the LSS. TNG300 contains $2500^3$ dark matter particles, with a baryonic mass resolution of $1.1\times 10^7 \rm \ M_{\odot}$, a comoving value of the adaptive gas gravitational softening length for gas cells of $370$ comoving parsec, gravitational softening of the collisionless component of $1.48$ kpc, and dark matter mass resolution of $5.9\times 10^7\ \rm M_{\odot}$. The TNG simulations adopt the \citetalias{ade2016planck} cosmological parameters. LC-TNGX is constructed with the box remap technique~\citep{carlson2010embedding}, and spans across redshifts $0.03 \lesssim z \lesssim 0.3$; this range was motivated by initial MW-mass scale of the hot CGM observations (e.g. \citealt{comparat2022erosita, Chadayammuri:2022us, zhang2024hot}). It goes out to 1231  comoving  Mpc (cMpc) along the x-axis, subtending an area of $47.28\rm \ deg^2$ on the sky in the y-z plane. There are 22 snapshots within the observationally motivated redshift range  of $0.03 \leq z \leq 0.3$. 

The physical properties of the distinct halos and subhalos within the TNG300 lightcone are obtained by the Friend-of-Friend (FoF)  and \textsc{subfind} algorithm~\citep{springel2001populating, dolag2009substructures}. \textsc{subfind} detects gravitationally bound substructures, equivalent to galaxies in observations, and also provides us with a classification of subhalos into centrals and satellites, where centrals are the most massive substructure within a distinct FoF halo. 

This paper focuses its analysis on studying the average XSB profiles from central galaxies in halo mass bins. Therefore, we use the FoF groups, whose centres are defined as the most bound particle within the central subhalo as found by the \textsc{subfind} algorithm. We do not consider the satellite galaxies as the average XSB profiles from satellites probe the X-ray emission of their more massive hosts; see \citet{shreeram2025quantifying} for further discussion and the quantification of this effect.

The X-ray photons are simulated within the LC-TNGX in the $0.5-2.0$ keV intrinsic band with \texttt{pyXsim}~\citep{zuhone2016pyxsim}, which is based on \textsc{phox}~\citep{Biffi:2013uh}, by assuming an input emission model where the hot X-ray emitting gas is in collisional ionization equilibrium. The X-ray emissivity, $\epsilon$, in turn depends on the gas mass density $\rho$, temperature $T$ and the metallicity of the gas $Z_{\rm met}$ as follows~\citep{Lovisari:2021aa}, 
\begin{equation}\label{eq:xray_emiss}
    \epsilon =  n_{\rm e} n_{\rm p} \Lambda(T, Z_{\rm met}),
\end{equation}
where $\rm n_e$ and $\rm n_p$ are the number densities of the electron and protons, which are related to the gas mass density $\rho = \mu m_{\rm p} (n_{\rm e} + n_{\rm p})$. Here, $\mu$ and $m_{\rm p}$ are the mean molecular weight and the proton mass, respectively. $\Lambda(T, Z_{\rm met})$ is the cooling function of the hot gas, which depends on the emission  mechanism in the considered energy window\footnote{these could be free-free, recombination, or line cooling; see \cite{bohringer2010x} for a review.}.  The spectral model computations of hot plasma use the Astrophysical Plasma Emission Code, \textsc{apec}\footnote{APEC link \url{https://heasarc.gsfc.nasa.gov/xanadu/xspec/manual/node134.html}} code~\citep{smith2001collisional} with atomic data from \textsc{atomdb} v3.0.9~\citep{foster2012updated} and solar abundance values from~\cite{anders1989abundances}. This model uses the plasma temperature of the gas cells (in keV), the redshift $z$ and metallicity. We have improved the mock X-ray generation from \cite{shreeram2025quantifying}, where we assumed a constant metallicity of $0.3\ Z_\odot$, to include the intrinsic TNG gas cell metallicity. This is a key improvement as it allows a more accurate and self-consistent estimate of the X-ray emission at the gas cell level throughout the lightcone. 

The events are generated by assuming a telescope with an energy-independent collecting area of $1000$ cm$^2$ and an exposure time of $1000$ ks. The photon list is generated in the observed frame of the X-ray emitting gas cells and is corrected to rest frame energies. Finally, the photons generated by the gas cells are projected onto the sky. We use the projected central galaxy positions and obtain the XSB profiles in the $0.5-2.0$ keV band. We select the X-ray events within $R_{\rm 200m}$\footnote{$R_{\rm 200m}$ and $M_{\rm 200m}$ is the radius and mass at which the density of the halo is  $200\times$ the mean matter density (cold dark matter and baryons).} of the parent halo for obtaining the XSB profiles and we define these profiles as the intrinsic hot gas emission profiles. In this work, we test the impact of the CGM environment on the XSB profiles. We show the distribution of X-ray events in the bottom panel of Fig.~\ref{fig:lss}. For this purpose, we classify the cosmic web in the LC-TNGX into different LSEs, as further described in the following sections.

\subsection{Extracting cosmic filaments in LC-TNGX using DisPerSE}
\label{subsec:disperse}
The theoretical background of DisPerSE is provided in \cite{sousbie2011persistenta, sousbie2011persistentb}\footnote{\url{http://www2.iap.fr/users/sousbie/web/html/indexd41d.html}}; here we summarise the most important details. DisPerSE deals with discrete datasets and provides the user with a geometric three-dimensional ridge, allowing for the classification of the cosmic web based on its topology. It is built on Morse and persistence theories, and it functions by first estimating the underlying density field, given an input galaxy distribution, using the Delaunay tessellation field estimator (DTFE; \citealt{schaap2000continuous, cautun2011dtfe}). The Delaunay tessellation is a triangulated space representing a geometric assembly of cells, faces, edges and vertices mapping the entire volume of the galaxy distribution. The gradients of the DTFE density field, $\rho_{\rm DTFE}$, provide the critical points: maxima, minima and saddles, which are connected by field lines tangent to the gradient of $\rho_{\rm DTFE}$. The filaments comprise connecting segments between the maxima of the density field, also called CPmax or peaks, to the saddle points.

The significance of a filament, or the persistence threshold, is estimated by the density contrast of the critical pair chosen to pass a certain signal-to-noise threshold. For a filament, this critical pair is between a CPmax and the saddle point. The noise level is defined relative to the root mean square of the significance values obtained from random sets of points. This thresholding eliminates less significant filaments, simplifying the Morse complex and retaining its most topologically robust features. 
The skeletons generated in this work use the $3\sigma$ persistence thresholds, following the careful calibration method presented in \cite{Galarraga-Espinosa:2024ab}. The galaxies above $M_\star > 10^9\ \MSUN$ ($83,297$ galaxies) were used to build the skeleton, shown in Fig.~\ref{fig:lss}. This mass choice is motivated to match typical galaxy mass limits in current observation surveys, and the resolution of the simulation used to build the lightcone. We highlight that while the filament skeleton is constructed from DisPerSE, using the galaxy distribution, as shown by the top panel of Fig.~\ref{fig:lss}, we find that the hot gas, as probed by the X-rays, also traces the cosmic web as shown in the bottom panel of Fig.~\ref{fig:lss}. This alignment provides a compelling validation of the full pipeline used in this work, from the construction of the lightcone and the generation of X-ray maps to the identification of the filamentary network.

 \begin{figure}
    \centering
    \includegraphics[width=\columnwidth]{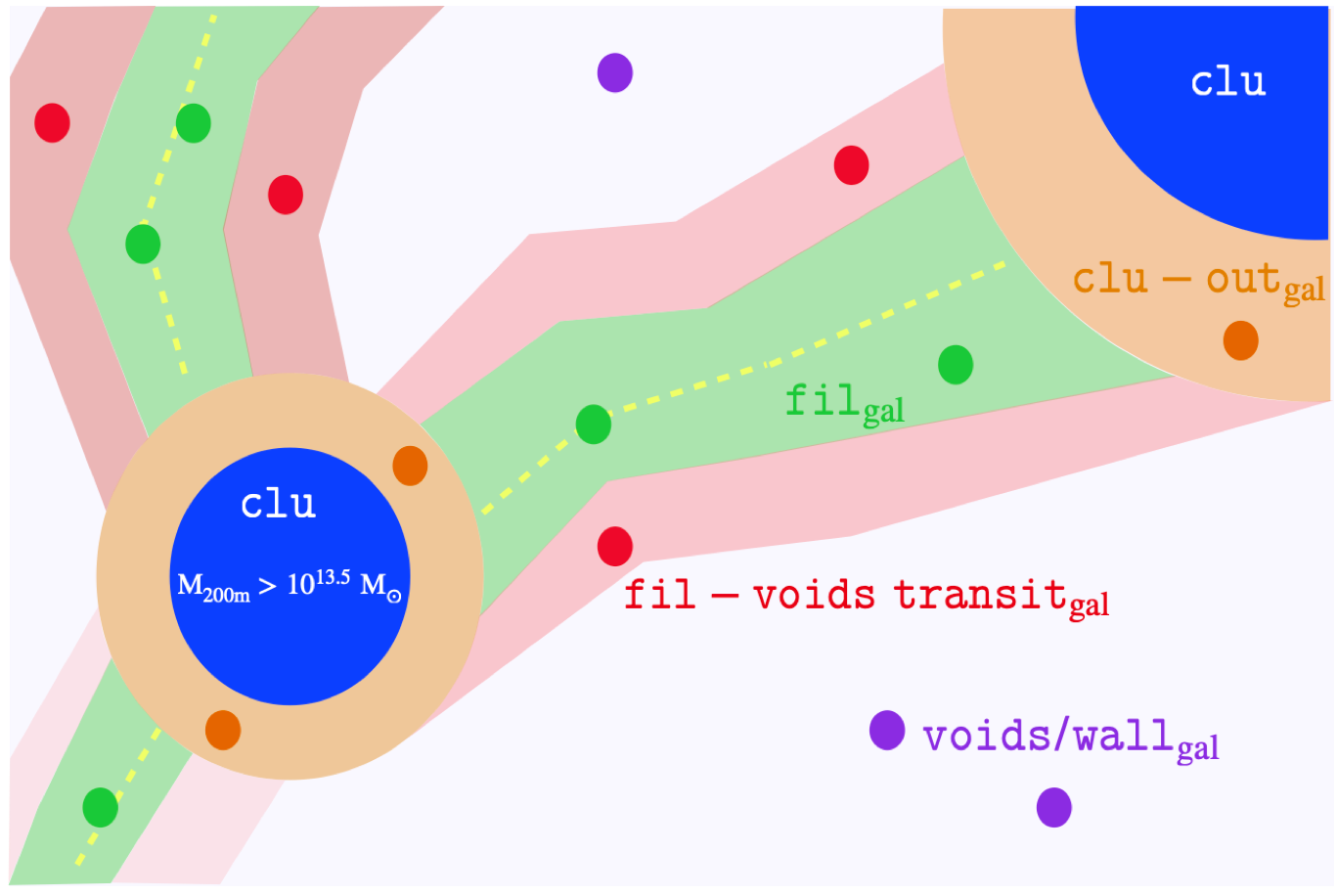}
    \caption{Illustration of the classification of galaxies into different LSE categories within the LC-TNX. The filament spines (yellow dashed lines) are extracted using the DisPerSe algorithm. The clusters, representing the nodes of the cosmic web, are defined as halos with masses $\Mtwo>10^{13.5}\ \MSUN$. The dots represent central galaxies in different LSE, such as cluster outskirts (orange), filaments (green), filament-void transition region (red), and voids/walls (purple).}
    \label{fig:lss_illustration}
\end{figure}

\begin{table*}[]
\centering
\caption{Fraction of galaxies in each halo mass bin for each large-scale environment classification. }
\label{tab:classification}
\resizebox{\textwidth}{!}{%
\begin{tabular}{llccccc}
Color & LSS classifications & \begin{tabular}[c]{@{}c@{}}$\rm M_{200m} \in 10^{11.5-12} M_\odot$\\ $28, 492$ halos\end{tabular} & \begin{tabular}[c]{@{}c@{}}$\rm M_{200m} \in 10^{12-12.5} M_\odot$\\ $11, 094$ halos\end{tabular} & \begin{tabular}[c]{@{}c@{}}$\rm M_{200m} \in 10^{12.5-13} M_\odot$\\ $3, 677$ halos\end{tabular} & \begin{tabular}[c]{@{}c@{}}$\rm M_{200m} \in 10^{13-13.5} M_\odot$\\ $1, 113$ halos\end{tabular}& \begin{tabular}[c]{@{}c@{}}All $\rm M_{200m} \in 10^{11.5-13.5} M_\odot$\\ $44, 376$ halos\end{tabular}\\ \hline \\
\textcolor[HTML]{ff7c00}{$\blacksquare$} & Cluster outskirts & $3\% \ (975)$ & $3\%\ (304)$ & $3\%\ (113)$ & $2\% \ (25)$ & $3\%\ (1417)$\\
\textcolor[HTML]{1ac938}{$\blacksquare$} & Filaments & $12\%\ (3,470)$ & $10\% \ (1,139)$ & $13\% \ (465)$ & $20\% \ (221)$ & $12\%\ (5,295)$ \\
\textcolor[HTML]{e8000b}{$\blacksquare$} & Filament-void transition & $23\% \ (6,627)$ & $22\% \ (2,391)$ & $22\% \ (801)$ & $22\%\ (242)$ & $23\% \ (10,061)$ \\
\textcolor[HTML]{8b2be2}{$\blacksquare$} & Voids/Walls & $61\%\ (17,420)$ & $65\%\ (7,260)$ & $63\%\ (2,298)$ & $56\%\ (625)$ & $62\%\ (27,603)$\\
\end{tabular}%
}
\tablefoot{For every halo mass bin, we present the percentage of objects in cluster outskirts (orange), filaments (green), filament-void transition region (red), and voids/walls (purple) within LC-TNGX (detailed in Sec.~\ref {subsec:classification}).  The $M_{\rm 200m} > 10^{13.5} M_\odot$ halos, of which there are $551$, are defined as massive groups and clusters.}
\end{table*}

\begin{figure*}
    \centering
    \includegraphics[width=\textwidth]{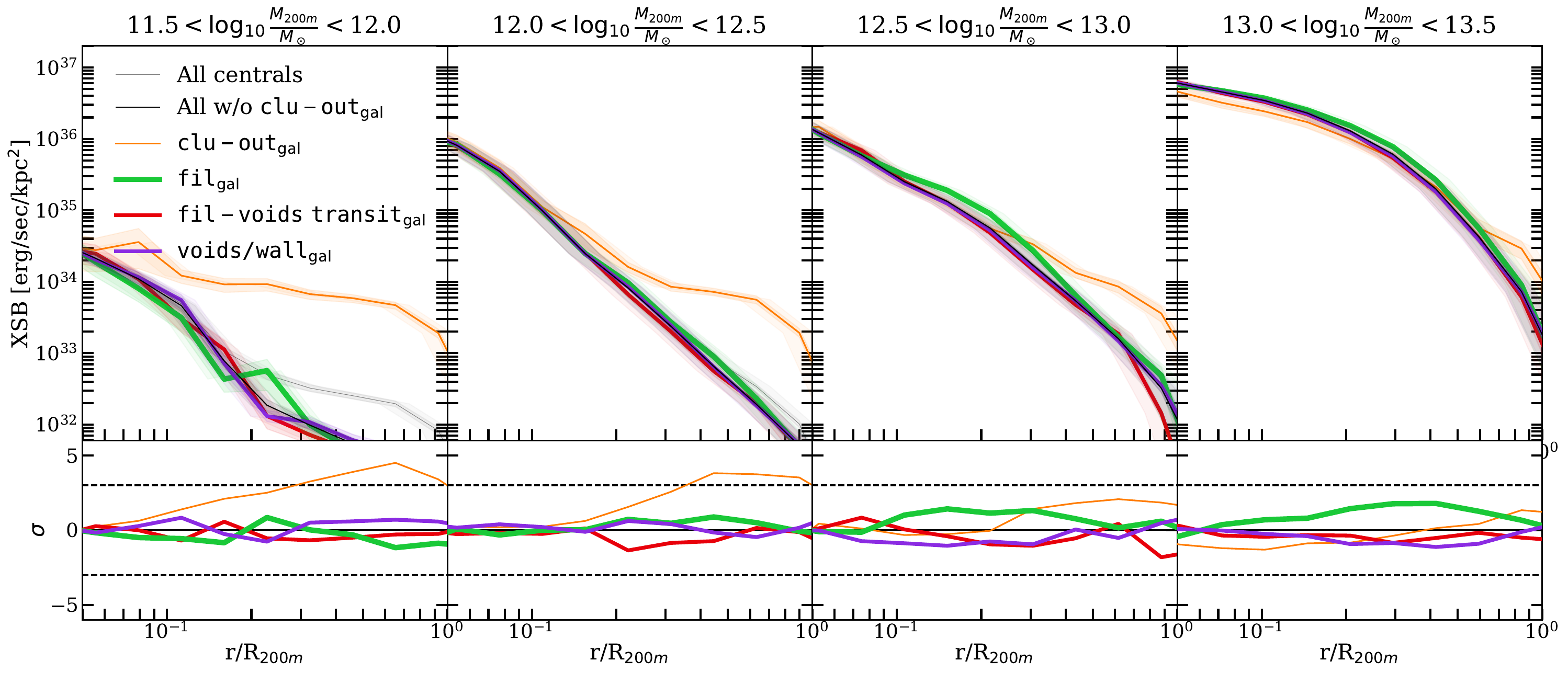}
    \caption{The X-ray surface brightness profiles (XSB) of halos located in different cosmic web environments in increasing halo mass bins from left to right. The fraction of central galaxies in each mass bin (and each LSE category) is shown in Tab.~\ref{tab:classification}. The bottom panel shows the significance (see Eq.~\ref{eq:dev}) of the difference in the XSB in a given cosmic environment with respect to the others (see text for details). The shaded regions around each curve correspond to the uncertainty in the mean profile obtained by bootstrapping.}
    \label{fig:Sx_halo_mass_bins}
\end{figure*}

\subsection{Classification on halos in LC-TNGX into different cosmic web environments}
\label{subsec:classification}
Given the skeleton from DisPerSE, the central galaxies in the LC-TNGX are divided into five mutually exclusive categories in a similar fashion as in \cite{Galarraga-Espinosa:2023aa}, also depicted in Fig.~\ref{fig:lss}. We summarise the fraction of galaxies in each of these distinct LSE categories in Tab.~\ref{tab:classification}, and we also illustrate the LSE classification of galaxies in Fig.~\ref{fig:lss_illustration}, which are defined as follows.

\begin{enumerate}
    \item Clusters and massive groups: The galaxy clusters and massive groups are defined as halos with $\Mtwo> 10^{13.5}\ \MSUN$, with a radius $\Rtwo$, centred on the positions of the FoF halos.

    \item Galaxies in cluster and group outskirts (\texttt{clu-out$_{\rm gal}$}): The category comprises of galaxies located $(1-3)\times \Rtwo$ from a cluster center. This radial range is motivated by \cite{Aung2023outskirt}, who demonstrated that the gas accretion shock is located at approximately $(1.5-3)\times \Rtwo$. This accretion shock leads to the onset of the galaxy quenched fraction, influenced by ram pressure and tidal stripping due to clusters, approaching the average value (e.g., \citealt{cen2014gas}). 
    
    \item Galaxies in filaments (\texttt{fil$_{\rm gal}$}): The cosmic filaments that are extracted by DisPerSe are defined as cylinders, with a radius of $1$ cMpc~\citep{wang2024boundary}, aligned along the spine of the filament skeleton identified by DisPerSe (Sec.~\ref{subsec:disperse}). The central galaxies that are within the $1$ cMpc cylinder of the filament spine are defined as galaxies in filaments.
    
    \item Galaxies in the filament-voids/walls transition region (\texttt{fil-voids\ transit$_{\rm gal}$}): The central galaxies that are $1-3$ cMpc away from the filament spine are categorised as galaxies in the passage between filaments and voids. This category is primarily constructed to define a smooth transition between the galaxies located in filament outskirts to those in voids. For brevity, we also call this group of galaxies the transition galaxies.
    
    \item Galaxies in voids and walls  (\texttt{voids/wall$_{\rm gal}$}): This category encompasses all the galaxies that do not fall in any of the above categories. These galaxies are also called "field" galaxies. The galaxies in voids and walls are combined due to the lack of information about the void or wall size, which is required to associate galaxies in these distinct categories.
    
\end{enumerate}
We present the fraction of central galaxies in each of these distinct categories in Tab.~\ref{tab:classification}; within LC-TNGX, of the $44,376$ halos with $\Mtwo \in 10^{11.5-13.5} \MSUN$, $3\%$, $12\%$, $23\%$, and $62\%$ of the halos are located in cluster outskirts, filaments, filament-void transition region and voids/walls, respectively.

We compare our findings on the fraction of filament galaxies with other literature works reporting the fraction of galaxies in filaments. \cite{ganeshaiah2019cosmic} find that $63\%$ of the central galaxies are hosted by filaments; they use the EAGLE simulation and run the Bisous filament finder on the galaxy distribution. Using the \textsc{TheThreeHundred} project, \cite{kuchner2022inventory} find that $45\%$ of the galaxies in filaments are feeding clusters (independent of the distance to the cluster). \cite{Navdha2025MNRASrelation}, using the Millennium simulation, to show that $26\%$ of the galaxies reside in the cosmic web at a halo mass of $10^{11}\MSUN$, going up to $50\%$ at $10^{12.7}\MSUN$. However, these fractions include both central and satellite galaxies, and they define the cosmic web differently. We draw extreme caution when directly comparing these values with the $12\%$ of filament galaxies found in our work. As also highlighted by these works, the fraction of filament galaxies is affected by the following: (1) the box size and the volume of the simulation, in turn affecting the number of massive clusters in the cosmic web, which are different across the simulations discussed here; (2) different filament finding techniques (see e.g.,  \citealt{cautun2013nexus, leclercq2016comparing, Libeskind2018compare, rost2020comparison}, for comparisons between different cosmic web identification methods); and (3) different cosmic web definitions, affecting in turn the resulting fraction of galaxies in filaments. More so, we are for the first time performing such a categorisation of galaxies in a lightcone configuration (ranging across $0.03 \leq z\leq 0.3$), which could show further deviations from a galaxy categorisation in a cubic snapshot at a fixed redshift.

\begin{figure*}
    \centering
    \includegraphics[width=.95\textwidth]{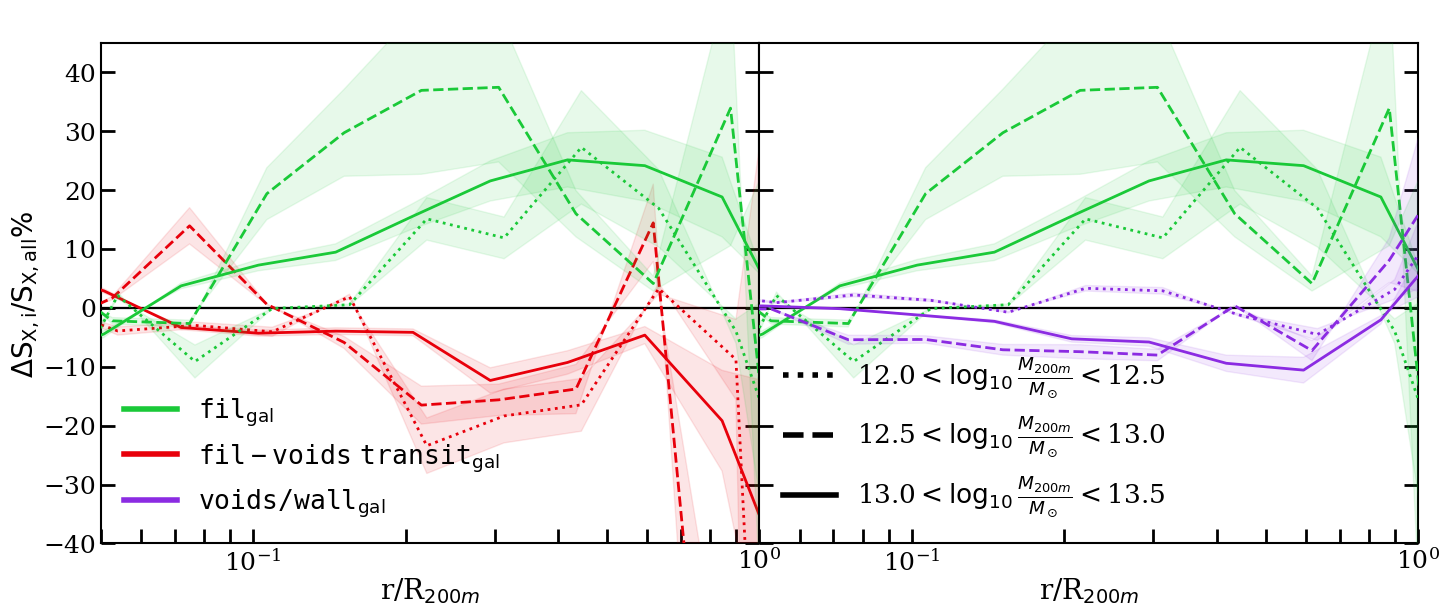}
    \caption{The percentage deviations of the XSB profiles of halos located in filaments (green), compared to those in the filament-void transition region (left panel; red lines) and in voids/walls (right; purple lines). We also show the effect of varying the halo mass bins, where the dotted lines represent the $\Mtwo \in 10^{12-12.5}\ \MSUN$ halos, whereas the dashed and solid lines represent the $\Mtwo \in 10^{12.5-13}\ \MSUN$ and $\Mtwo \in 10^{13-13.5}\ \MSUN$ bins, respectively. We find that the XSB profiles of \texttt{fil$_{\rm gal}$} are X-ray brighter in $\sim (0.3-0.5)\times \Rtwo$ by $20-45\%$ with respect to the the  \texttt{fil-voids\ transit$_{\rm gal}$} and \texttt{voids/wall$_{\rm gal}$} populations. }
    \label{fig:deviation}
\end{figure*}

\section{Effect of the environment on the X-ray surface brightness profiles of halos}
\label{sec:results}
The hot CGM emits in X-rays due to infalling gas within halos, which is shock-heated up to the virial temperatures, $T_{\rm vir} \gtrsim 5\times 10^{5}$ K~\citep{white1978core}. We present the projected X-ray surface brightness profiles of halos for varying halo masses and cosmic web environments in Fig.~\ref{fig:Sx_halo_mass_bins}. The increasing halo mass bins (left to right) result in brighter XSB profiles, given that the gas is heated to higher virial temperatures~(e.g., \citealt{comparat2022erosita, zhang2024hot}). 

\begin{figure*}
    \centering
    \includegraphics[width=0.9\textwidth]{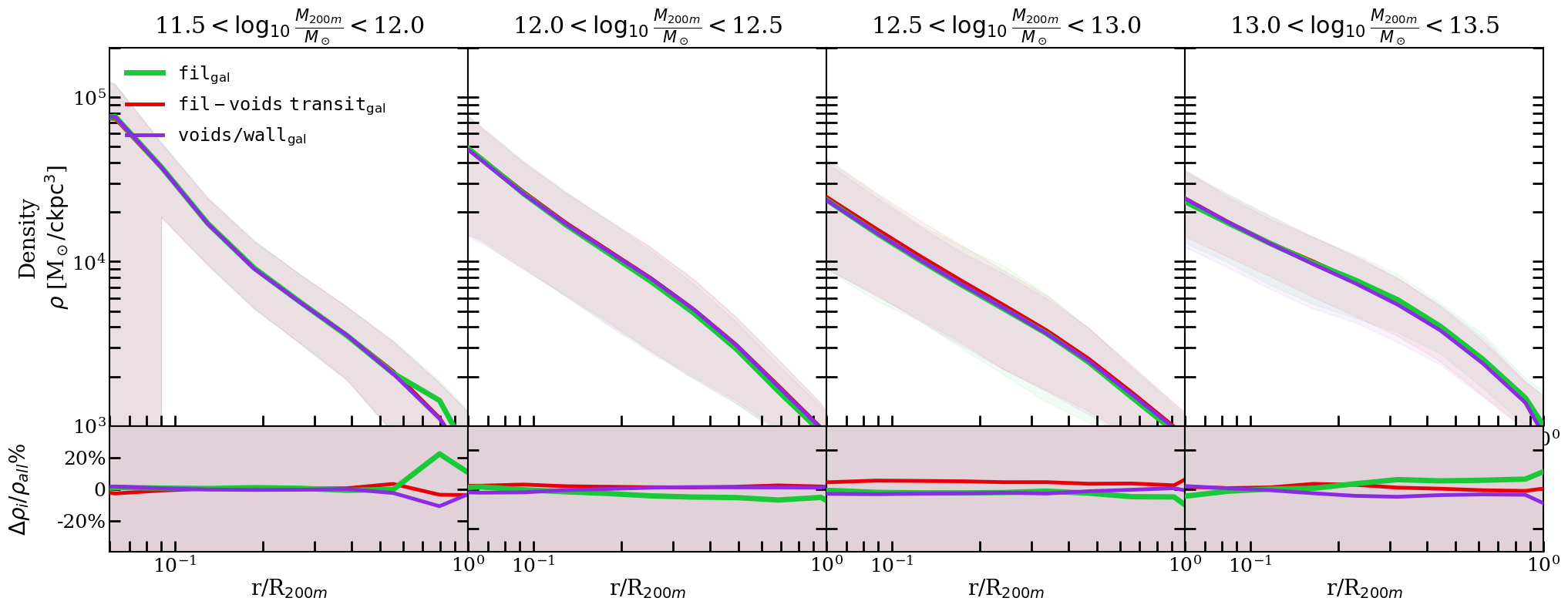}
    \includegraphics[width=0.9\textwidth]{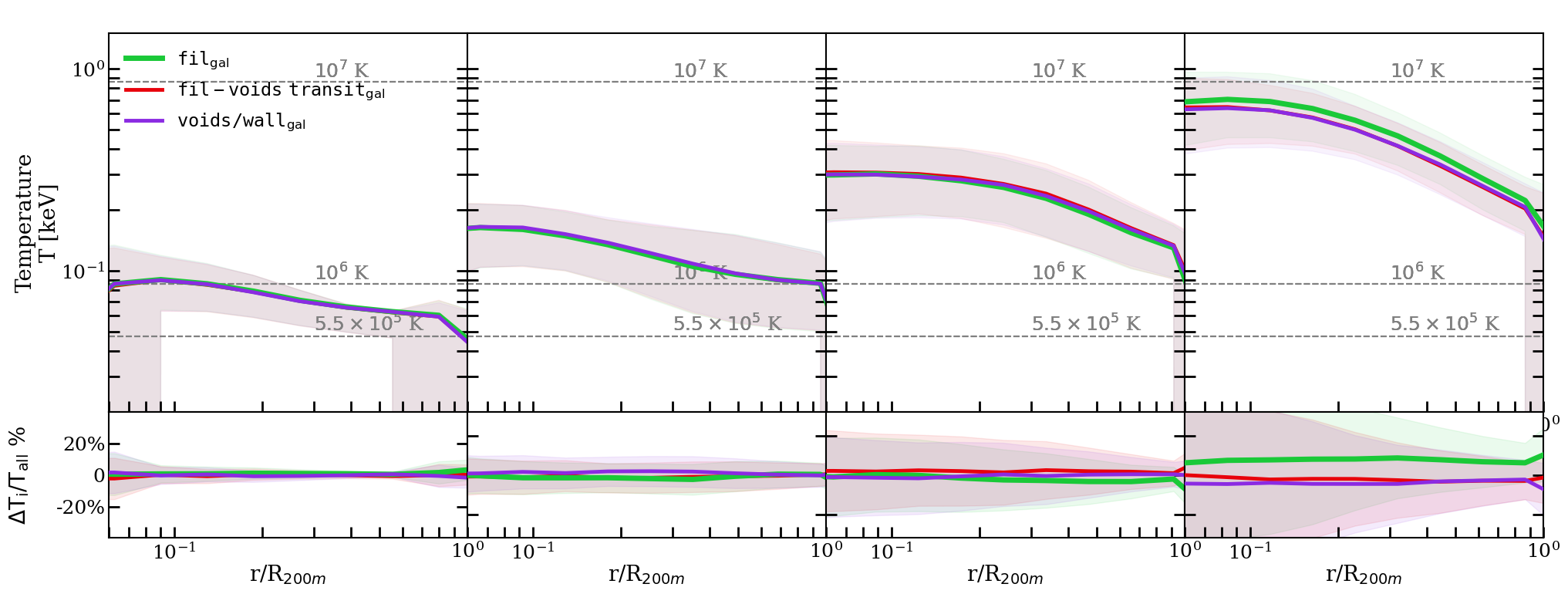}
    \includegraphics[width=0.9\textwidth]{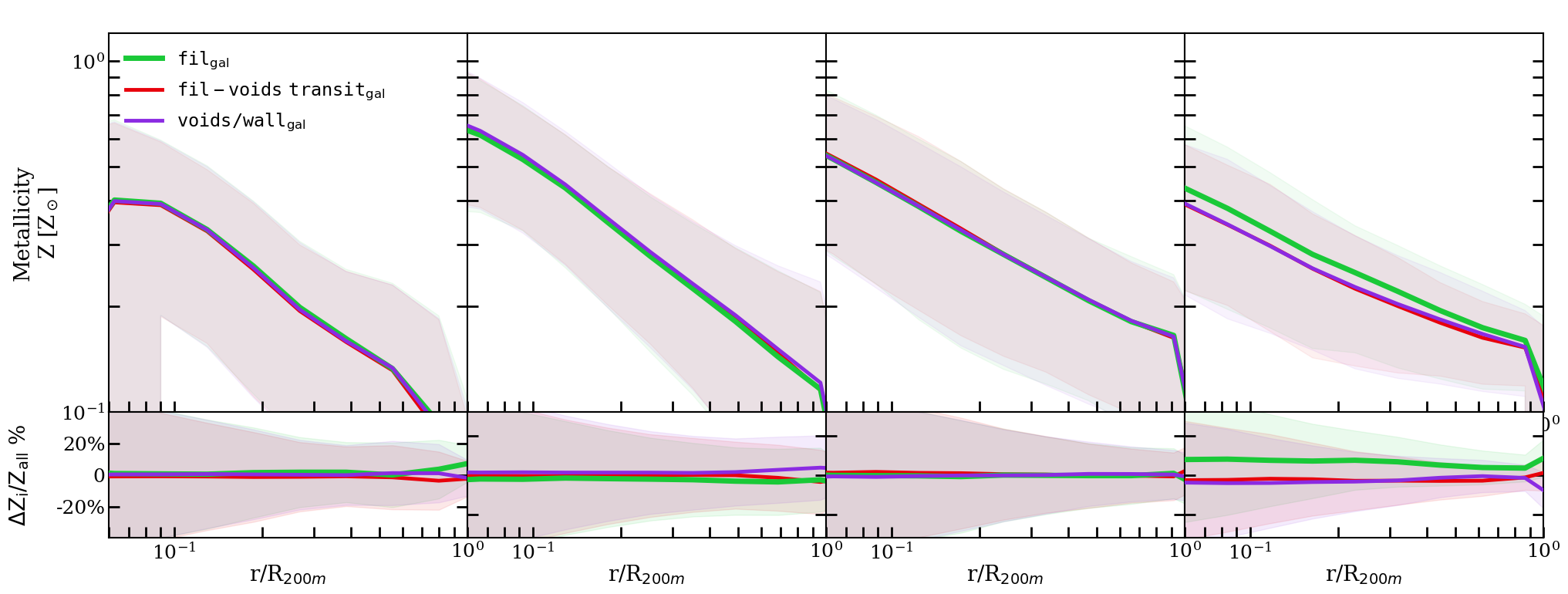}
    \caption{The volume weighted gas mass density (top panel), temperature (middle panel) and metallicity (bottom panels) profiles together with their percentage deviations (corresponding bottom rows) of halos located in different LSE in increasing halo mass bins from left to right. The different cosmic environments considered here are filaments (green), filament-void transition region (red), and voids/walls (purple). The shaded regions show the $16^{th}-84^{th}$ percentile distribution.}
    \label{fig:thermo}
\end{figure*}

We separate galaxies into the following different bins of halo mass: $\Mtwo \in 10^{11.5-12}\ \MSUN$, $\Mtwo \in 10^{12-12.5}\ \MSUN$, $\Mtwo \in 10^{12.5-13}\ \MSUN$ and $\Mtwo \in 10^{13-13.5}\ \MSUN$. These are hereafter referred to as the lowest, low, medium, and highest mass bins, respectively.
For the lowest halo mass bin (leftmost panel), the peak temperature of the gas within the halo is $0.03-0.09$ keV (also shown later in Fig.~\ref{fig:thermo}), which is well below the $0.5-2$ keV soft X-ray band considered here for measuring the XSB profiles. Therefore, the results of the XSB profiles corresponding to this lowest mass bin are only probing the high-temperature end of the lowest halo mass bin halos in the XSB profiles presented here, thereby explaining their weak signal.

The different LSE considered here are as follows, as also shown in Fig.~\ref{fig:Sx_halo_mass_bins} are cluster outskirts (orange), filaments (green), filament-void transition region (red) and voids/walls (purple). The bottom panel of Fig.~\ref{fig:Sx_halo_mass_bins} shows the significance of the deviation of a given XSB profile, $S_{X, i}$, in a defined LSE category with respect to all the other categories, $\Tilde{S}_{X, i}$. Here, the subscript $i$  signifies one among the five LSE categories defined in Sec.~\ref{subsec:classification} (also shown by the colored lines in Fig.~\ref{fig:Sx_halo_mass_bins}), while $\Tilde{S}_{X}$,{with a tilde}, is the XSB profile obtained by averaging over all the galaxies in the other LSE categories. The significance, $\sigma$, is defined as
\begin{equation}\label{eq:dev}
    \sigma = \frac{S_{X, i} - \Tilde{S}_{X}}{\sqrt{\Tilde{\delta}^2 + \delta_{i}^2} },
\end{equation}
where $\delta_{i}$ and $\Tilde{\delta}$ are the uncertainties in the mean XSB profiles obtained from bootstrapping. A positive significance implies that a given $S_{X, i}$ is brighter at a given scale compared to its counterparts in other LSE, while the negative significance represents steeper or X-ray fainter profiles.

The following sections detail the trends observed in the XSB profiles in the different LSE categories, where we present the effect of the LSE on the XSB profiles of halos in cluster outskirts in Sec.~\ref{subsubsec:clu_out} and in filaments and voids/walls in Sec.~\ref{subsubsec:fil_voids}.

\subsection{Galaxies in cluster outskirts}
\label{subsubsec:clu_out}

The \texttt{clu-out$_{\rm gal}$} galaxy population probes the XSB profile of the halos located physically close to a cluster, shown by the orange lines in Fig.~\ref{fig:Sx_halo_mass_bins}. Across all halo mass bins (left to right), although notably more prominent at the lowest and low halo mass bins, the XSB profiles from \texttt{clu-out$_{\rm gal}$} are significantly brighter than the mean XSB profile (without categorising halos by their LSE). We attribute this striking feature of XSB profiles from \texttt{clu-out$_{\rm gal}$} to its close proximity to a cluster, where the cluster outskirt begins to dominate the profile at these radii. We further quantify this effect in the significance plots in the bottom panel of Fig.~\ref{fig:Sx_halo_mass_bins}. We show that the halos in the least-massive bin are $3\sigma$ brighter at $0.2\ \Rtwo$ and $5\sigma$ brighter at $0.7\ \Rtwo$. The significance reduces to $3\sigma$ at $0.7\ \Rtwo$ for the low halo mass bin. For the medium and highest halo mass bins considered here, the significance of the XSB profiles from \texttt{clu-out$_{\rm gal}$} being brighter than their counterparts in other LSE is $<3\sigma$ significant at all radii. We explain the lower significance of the highest halo mass bins by the fact that more massive halos have comparable intrinsic X-ray brightness to their neighbouring cluster. Therefore, the cluster emission only dominates at large radii. Our results show that the X-ray contamination from the cluster is an important effect to consider when studying the galaxies in the cluster outskirts in X-ray stacking experiments.

Cluster and group outskirts are special regions, where, due to gas stripping processes, the gas in these halos is depleted, as shown by simulations (e.g.,~\citealt{bahe2013does, cen2014gas}) and observations  (e.g., \citealt{tanaka2004environmental, catinella2013galex, cortese2011effect}). However, in order to measure the impact of such stripping processes on the average XSB profiles obtained in stacking experiments, one must accurately model out the cluster XSB contribution. We leave such a modelling analysis to recover the XSB of galaxies in cluster outskirts from the contamination of the nearby cluster emission for future work. The following sections focus on the other LSE categories that are unaffected by cluster emission, as they are located far ($>3\Rtwo$) from the cluster and group centres.

\begin{figure*}
    \centering
    \includegraphics[width=\textwidth]{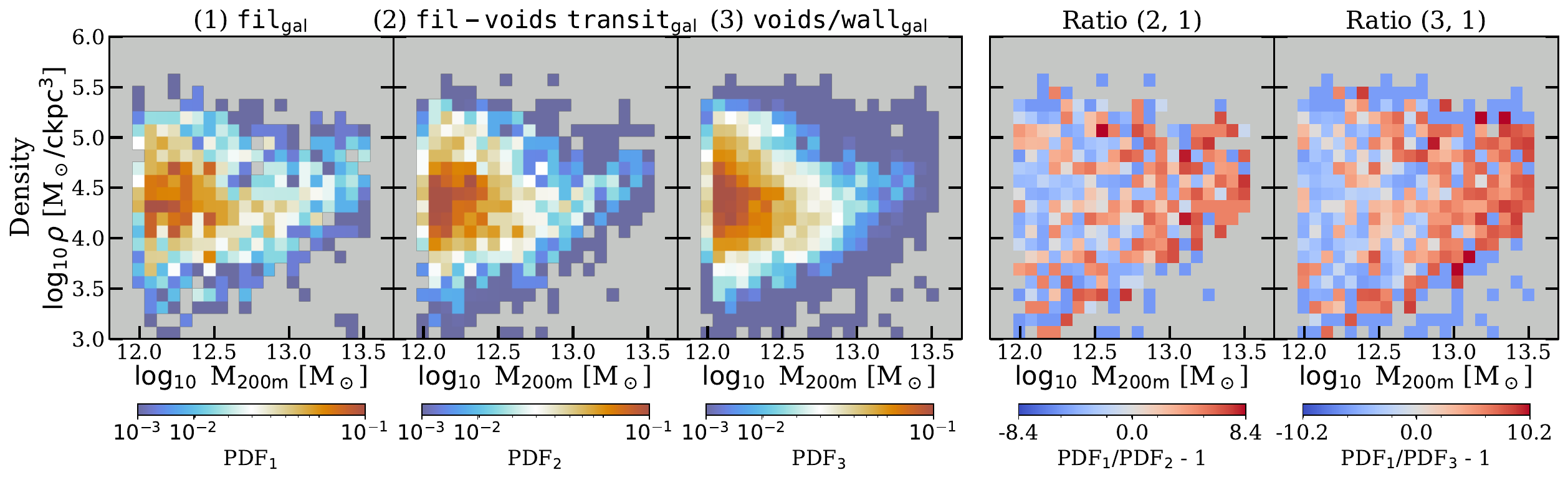}
    \includegraphics[width=\textwidth]{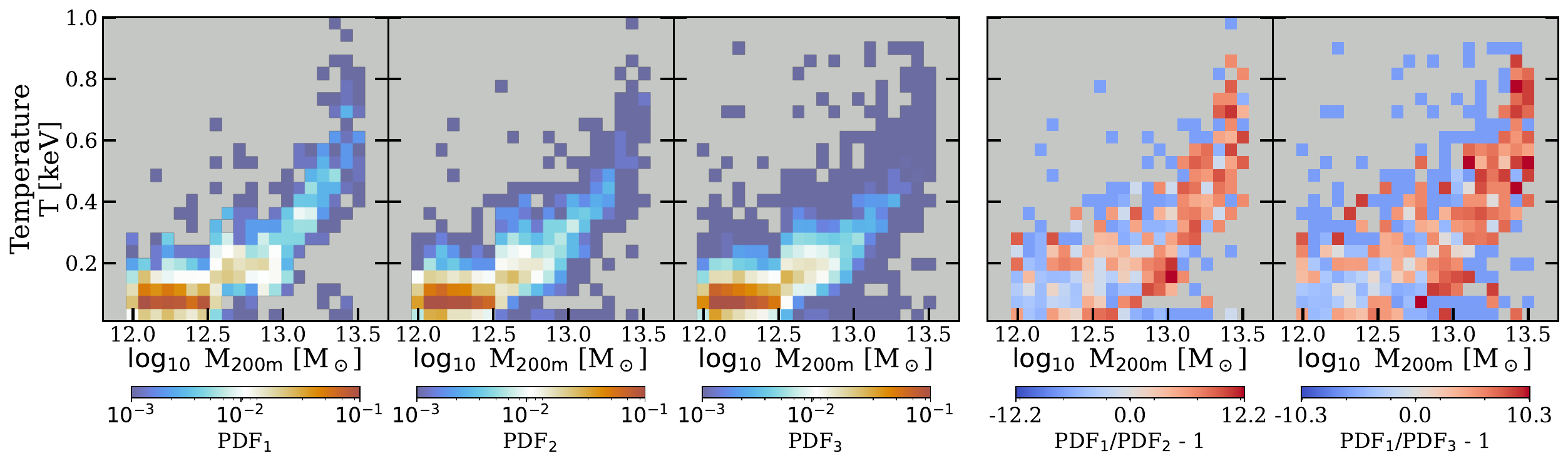}
    \includegraphics[width=\textwidth]{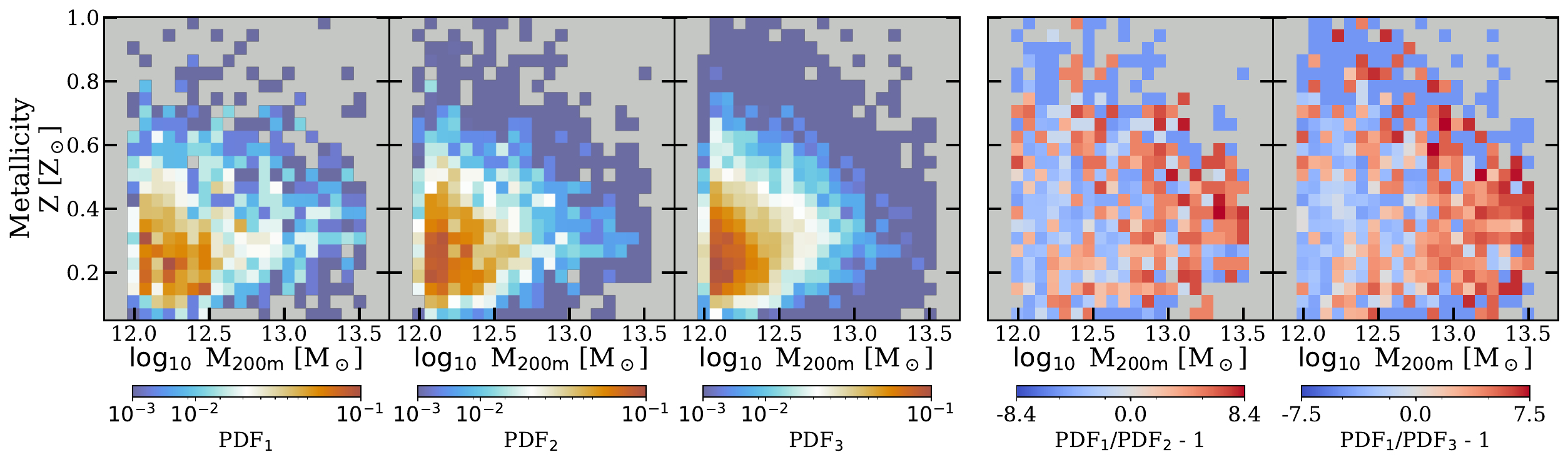}
    \caption{The normalised probability distribution functions of gas mass density, $\rho$ (top row), temperature, $T$ (middle row), metallicity $Z_{\rm met}$ (bottom row) as a function of the halo mass. The population of galaxies in filaments ($1,825$), filament-void transition region ($3,434$), and voids/walls ($10,183$) is shown in the first three columns, respectively. The last two columns show the ratios of the \texttt{fil-voids\ transit$_{\rm gal}$}  and the \texttt{voids/wall$_{\rm gal}$} to the \texttt{fil$_{\rm gal}$} population. The colour bars in the ratio panels indicate the relative suppression (blue) or enhancement (red) of the thermodynamic quantity of the filament galaxies with respect to transition or voids/walls galaxies. We find that the overdense intrafilamentary environments hosting the \texttt{fil$_{\rm gal}$} population show up to a factor of $10\times$ higher gas densities, $12\times$ higher temperatures, and $8\times$ higher metallicities compared to both the transitional and voids/walls populations. }
    \label{fig:interpretation}
\end{figure*}

\subsection{Galaxies in filaments and voids}
\label{subsubsec:fil_voids}
We now focus on the galaxies far away from clusters and massive groups, which are located in filaments, voids/walls and the transition region in between them. The \texttt{fil$_{\rm gal}$}, \texttt{fil-voids\ transit$_{\rm gal}$}, and \texttt{voids/wall$_{\rm gal}$} population are shown by the green, red, and purple lines in Fig.~\ref{fig:Sx_halo_mass_bins}, respectively, with the bottom panel showing the significance of deviation compared to other LSE. We further study these three categories in  Fig.~\ref{fig:deviation}, showing the absolute deviation in the XSB profiles in the low, medium and highest halo mass bins. We exclude the lowest halo mass bin, $\Mtwo \in 10^{11.5-12}\ \MSUN$, in this comparison as there is no observable trend due to the high scatter in the measured XSB profiles. We quantify the deviation of the filament galaxies with respect to transition galaxies (left panel), and the voids/walls galaxies (right panel). 

We find that for all the halo mass bins considered, the XSB profiles of \texttt{fil$_{\rm gal}$} are X-ray brighter than the mean population between $\sim (0.3-0.5)\times \Rtwo$ by $20-45\%$. More precisely, Fig.~\ref{fig:deviation} shows that the filament galaxies in the low, medium and highest halo mass bins are brighter than the mean XSB profile at different fractions of $\Rtwo$. For the low halo mass bin, the maximum mean deviation is $20\%$ at $\sim 0.45\times \Rtwo$, whereas for medium and highest halo mass bins, the maximum mean deviation of $40\%$ and $25\%$ is at $\sim 0.3 \Rtwo$ and $\sim 0.5 \Rtwo$, respectively. 

Interestingly, we also find that the voids/walls and the transition galaxies are X-ray fainter than the mean XSB profiles. In the \texttt{fil-voids\ transit$_{\rm gal}$} case, for all halo mass bins, the maximum mean deviation is between $-(10-20)\%$ at $\sim (0.2-0.3)\times \Rtwo$. This trend becomes weaker for the \texttt{voids/wall$_{\rm gal}$} populations, where all the XSB profiles are fainter than the mean profiles by less than $10\%$ across all halo mass bins.

We delve deeper into understanding the non-negligible XSB excess in the filamentary galaxies with respect to those in voids/walls and the transition galaxies by exploring the primary quantities affecting the X-ray emission, which are the underlying thermodynamic quantities, as further detailed in the following section.

\section{Discussion on filament galaxies being X-ray brighter than the galaxies in voids/walls}
\label{sec:discussion}

We investigate the thermodynamic properties, such as density, $\rho$, temperature, $T$, and metallicity, $Z_{\rm met}$, of the hot gas ($T>5\times 10^5$ K)\footnote{To ensure that we only include gas cells that physically emit in X-rays, we also exclude star-forming gas cells, and gas cells with densities above $10^{-25}$ g/cm$^3$~(see \citealt[Sec. 3.1]{shreeram2025quantifying} and \citealt[Appendix B-C]{truong2020x}).} around galaxies given the dependence of X-ray emission on the gas properties, as shown in Eq.~\ref{eq:xray_emiss}. In Fig.~\ref{fig:thermo}, we present the radial profiles of $\rho$ (top panel), $T$ (middle panel) and $Z_{\rm met}$ (bottom panel) for galaxies in filaments, voids/walls and the transitional region for increasing halo mass bins (left to right columns). The thermodynamic profiles shown in Fig.~\ref{fig:thermo} are computed using gas cell information combined in a volume-weighted manner in 3D. The volume-weighting accounts for the cell refinement criterion of the simulation. Indeed, the moving-mesh \textsc{arepo} code is based on a fixed mass threshold for the gas cells~\citep{weinberger2020arepo}, leading to a broad distribution of cell volumes. Therefore, our volume-weighted results are independent of the irregular volumes of the Voronoi gas cells in the \textsc{arepo} grid. Contrary to the more conventional mass-weighted average profiles, which are biased by the diﬀerent gas cell volumes.

For the highest halo mass bin in Fig.~\ref{fig:thermo}, we find that the mean $T$ and $Z_{\rm met}$ radial profiles are approximately $10\%$ higher in filament galaxies compared to those in voids/walls, and the transition population. The mean density profiles also show an enhancement at $r > 0.2\ \Rtwo$. However, due to the significant scatter around the mean profiles across different environments (as also illustrated in Fig.~5 of \citealt{Truong2023MNRASline}), these average trends alone are insufficient to robustly interpret the observed X-ray enhancement. To better explore the full variations across galaxy populations in different LSEs, we study the probability distribution functions (PDFs) of the average thermodynamic properties of individual galaxies as a function of the halo mass. 

Using the radial profiles presented in Fig.~\ref{fig:thermo}, we compute the average value of $\rho$, $T$, and $Z_{\rm met}$ for each galaxy within $\Rtwo$. In Fig.~\ref{fig:interpretation}, we present the resulting normalised PDFs of these average thermodynamic quantities as a function of halo mass. The top, middle, and bottom panels correspond to the two dimentional PDFs of $\rho$--$M_{\rm 200m}$, $T$--$M_{\rm 200m}$, and $Z_{\rm met}$--$M_{\rm 200m}$, respectively. The first three columns present the normalised PDFs for galaxies residing in filaments (left), the filament-void transition region (middle), and voids/walls (right). The last two columns show the fractional difference in the PDFs relative to the filament population: column four compares the transition population to filaments, and column five compares the voids/walls to filaments. The colour bars in the ratio panels indicate the relative suppression (blue) or enhancement (red) with respect to the filament galaxies. Across all three thermodynamic quantities, filament galaxies exhibit systematic enhancements compared to other large-scale environments. More specifically, galaxies in filaments are up to a factor of $10\times$ more likely to have higher gas densities, $12\times$ more likely to have higher temperatures, and $8\times$ more likely to have higher metallicities than galaxies in transition regions or voids/walls. These differences are especially pronounced in the medium-to-highest halo mass range ($10^{12.5}\ \MSUN < \Mtwo < 10^{13.5}\ \MSUN$), where the contrast between environments is strongest. Within this mass bin, the PDF ratios between filament galaxies and those in voids/walls or transition regions are, on average, higher by factors of $50{-}70 \%$ across all thermodynamic quantities. This comparative study of the gas properties that directly affect X-ray emissivity is crucial to better understand the results of Fig.~\ref{fig:deviation}. 

In the following, we briefly discuss other quantities that may impact X-ray emission. Another possibility for the \texttt{fil$_{\rm gal}$} population being X-ray brighter could be higher hot gas fractions. The importance of higher hot gas fractions on the X-ray brightness (and their detectability) of galaxy groups ($> 10^{13}\ \MSUN$) in Magneticum is discussed in \cite{marini2025impact}, where they show that X-ray bright groups are driven by higher hot gas fractions, alongside a steady accretion history and are located in overdense environments. Future work could jointly explore the impact of the assembly history on the hot gas fractions of the galaxies in different environments.

Finally, another possibility for enhanced X-ray emission from the \texttt{fil$_{\rm gal}$} population, given their denser environment than void galaxies, could be from gas clumping~\citep{Nagai2011clumping}\footnote{While \cite{Nagai2011clumping} demonstrate gas clumping to be important in cluster outskirts, this could also play a role for intrafilamentary environments.} or from penetrating gas streams~\citep{van2012properties, zinger2016role}. Particularly, the energy that is carried by infalling hot gas is dissipated and eventually radiated in X-rays at $r \gtrsim 70$ kpc (e.g., see Fig. 14 in \citealt{nuza2014distribution}); thereby affecting the XSB in the low-redshift universe. Another way to probe the past and present gas accretion activity of a galaxy can be tested using galaxy connectivity as a proxy~\citep{kraljic2020impact}.  Precisely, \cite{Galarraga-Espinosa:2023aa} demonstrate that galaxy connectivity is impacted by the large-scale environment, galaxy mass, and the local density, which they show to impact the star-formation rate of the galaxy. A future work which studies the contributions to the X-ray emission from infalling streams, as probed by galaxy connectivity, gas accretion rate, and gas clumping for the filament vs void galaxies, could illuminate whether these play an additional role in enhancing the X-ray emission in filament galaxies.

\section{Conclusions and summary}
\label{sec:conclusions}
In this work, we use an IllustrisTNG-based lightcone, LC-TNGX, to study the impact of the LSE on the hot gas properties of galaxies. We self-consistently generate mock X-ray observations within the LC-TNGX~\citep{shreeram2025quantifying}. We apply DisPerSe on the galaxy distribution to identify the cosmic filaments within LC-TNGX and thereby classify the central galaxies into different LSE. These are as follows: galaxies in cluster and group outskirts (\texttt{clu-out$_{\rm gal}$}), galaxies in filaments (\texttt{fil$_{\rm gal}$}), galaxies in the filament-void transition region (\texttt{fil-voids\ transit$_{\rm gal}$}), and galaxies in voids and walls  (\texttt{voids/wall$_{\rm gal}$}). We also study the effect of LSE in different halo mass bins. The main findings of this work are summarised below. 
\begin{enumerate}

    \item We show that the galaxies in cluster-outskirts are $\gtrsim 3\sigma$ brighter than the other populations at $\gtrsim 0.7 \Rtwo$ at lower halo masses ($\Mtwo < 10^{12.5}\ \MSUN$). Although this trend is $\lesssim 3\sigma$ significant for higher halo masses ($\Mtwo > 10^{12.5}\MSUN$), we find that the trend of cluster outskirts being brighter than their counterparts at large radii persists (Fig.~\ref{fig:Sx_halo_mass_bins}). This striking feature of XSB profiles from galaxies in cluster outskirts being brighter is attributed to their close proximity to a cluster, where the cluster outskirt begins to dominate the profile at these radii. We highlight the importance of this effect when studying the population of halos near clusters in X-ray stacking experiments.

    \item We find that the filament galaxies are X-ray brighter than the galaxies in voids/walls and in the transitional region between them. More precisely, independent of the halo mass bins considered here, the XSB profiles of filament galaxies are X-ray brighter between $\sim (0.3-0.5)\times \Rtwo$ by $20-45\%$ (Fig.~\ref{fig:deviation}). We investigate the source of this brightness by exploring the thermodynamic properties ($\rho$, $T$, and $Z$) of the hot gas in these galaxies in these different environments. We find that the filament galaxies show significantly enhanced thermodynamic properties compared to those in transition regions or voids/walls. They are up to $10\times$ more likely to exhibit higher gas densities, $12\times$ more likely to have higher temperatures, and $8\times$ more likely to show elevated metallicities. These differences peak in the $10^{12.5}\ \MSUN < \Mtwo < 10^{13.5}\ \MSUN$ range, where ratios of the probability density functions across all quantities are consistently $50{-}70 \%$ higher for filament galaxies.

\end{enumerate}

Our findings highlight the importance of environmental effects in interpreting the hot CGM in X-rays. The framework developed here opens several promising avenues for future work. On the observational side, with improved spectral resolution from ongoing/upcoming missions like \textit{XRISM}~\citep{tashiro2020status} and \textit{NewAthena} \citep{barret2020athena, cruise2025newathena}, and the already available depth of \textit{eROSITA} data \citep{merloni2024srg}, it is feasible to search for an X-ray excess in filament galaxies using stacking techniques (e.g., \citealt{zhang2024hot}) by including LSE classification with ongoing/upcoming stage-4 surveys (e.g. {Euclid}, {4MOST}, DESI, PFS). Additionally, future work should explore the energy dependence of the XSB profiles in the soft X-ray regime ($0.2$--$1$~keV), particularly for halos with $\Mtwo < 10^{12}\ \MSUN$, to better probe the impact of the LSE on lower mass halos in X-rays. On the theoretical side, future studies exploring the effects of gas clumping~\citep{Nagai2011clumping, Zhuravleva2013clumping, Avestruz2016stirr}, infalling hot streams~\citep{zinger2016role}, mass assembly history~\citep{marini2025impact}, galaxy connectivity~\citep{Galarraga-Espinosa:2023aa}, and impact of filament morphology~\citep{Galarraga-Espinosa:2024ab, yu2025impact} on filament galaxies could further elucidate the mechanisms driving enhanced X-ray emission in filaments. The work developed here lays the groundwork for jointly constraining CGM properties with the impacts of the cosmic web, key for understanding galaxy formation in a cosmological context.

\begin{acknowledgements}
SS would like to thank Fulvio Ferlito, Nabila Aghanim, R\"udiger Pakmor, and Annalisa Pillepich for the helpful scientific discussions.
This research was facilitated by the Munich Institute for Astro-, Particle and BioPhysics (MIAPbP), which is funded by the Deutsche Forschungsgemeinschaft (DFG, German Research Foundation) under Germany´s Excellence Strategy – EXC-2094 – 390783311.
IM acknowledges funding from the European Research Council (ERC) under the European Union's Horizon Europe research and innovation programme ERC CoG (Grant agreement No. 101045437).
Computations were performed on the HPC system Raven at the Max Planck Computing and Data Facility. We acknowledge the project support by the Max Planck Computing and Data Facility.
\end{acknowledgements}

\bibliographystyle{aa} 
\bibliography{biblio} 

\end{document}